\documentclass[aps,12pt,preprintnumbers]{revtex4}

\usepackage{graphicx} 
\usepackage{amsmath,amssymb,epsfig,bm,comment}

\renewcommand\d{\partial}
\newcommand\<{\langle}
\renewcommand\>{\rangle}
\newcommand\x{\mathbf{x}}
\newcommand\tr{\mathop{\mathrm{tr}}}
\renewcommand\L{\mathcal{L}}

\begin{document}

\preprint{INT-PUB-10-041}
\title{Deconstructing holographic liquids}
\author{Dominik Nickel and Dam T.~Son}
\affiliation{Institute for Nuclear Theory, University of Washington,
Seattle, Washington 98195-1550, USA}
\date{September 2010}

\begin{abstract}

We argue that there exist simple effective field theories describing
the long-distance dynamics of holographic liquids.  The degrees of
freedom responsible for the transport of charge and energy-momentum
are Goldstone modes.  These modes are coupled to a strongly coupled
infrared sector through emergent gauge and gravitational fields.  The
IR degrees of freedom are described holographically by the
near-horizon part of the metric, while the Goldstone bosons are
described by a field-theoretical Lagrangian.  In the cases where the
holographic dual involves a black hole, this picture allows for a
direct connection between the holographic prescription where currents
live on the boundary, and the membrane paradigm where currents live on
the horizon.  The zero-temperature sound mode in the D3-D7 system is
also re-analyzed and re-interpreted within this formalism.
%We suggest that the effective low-energy dynamics of
%theories dual to extremal black branes are also theories of Goldstone
%bosons coupled to degrees of freedom living in the AdS$_2$ infrared
%region.

\end{abstract}

\maketitle

\section{Introduction}
\label{sec:intro}

There is considerable interest in using holographic
methods~\cite{Maldacena:1997re,Gubser:1998bc,Witten:1998qj} to study
strongly coupled quantum liquids.  A typical example of such a liquid
is the ${\cal N}=4$ super-Yang-Mills plasma, which is frequently used
as a prototype for the strongly coupled quark gluon plasma created at
RHIC~\cite{Policastro:2001yc}.  The liquids are described
mathematically as a solution of a higher-dimensional theory in an
asymptotically AdS spacetime.  To compute correlations functions using
gauge/gravity duality, one solves field equations in the bulk with
boundary conditions at the AdS boundary.  The microscopic theory is
typically a large-$N$ gauge theory in the strong coupling limit.

Frequently, however, one is not interested in the details of the
microscopic theory, but only in the long-distance behavior at finite
temperature and/or density.  This is the regime relevant for the
hydrodynamic behavior at finite temperature and quantum critical
behaviors at zero temperature.  In fact, much of the recent
``AdS/CMT'' activities~\cite{Liu:2009dm,Cubrovic:2009ye} are directed
toward finding new quantum critical behaviors. One can then ask: what
are the minimal ingredients needed to describe the
long-distance dynamics of holographic liquids?  Is the full holographic
description needed?  As we will argue in this paper, it is not.  The
long-distance behavior of holographic liquids can be described by a
set of Goldstone bosons, interacting with a strongly coupled infrared
sector.  Holography may be needed to describe the infrared sector, but
not the Goldstone bosons.

That a strongly coupled infrared sector should appear in the
low-energy effective theory is rather clear.  According to the
dictionary of holography, low energies correspond to the near-horizon
part of the metric.  Various possible types of behavior of the metric
in the IR have been observed and
classified~\cite{Gubser:2008wz,Gubser:2009cg,Horowitz:2009ij}, and one
expects different IR asymptotics to correspond to different IR
sectors.  For example, a black hole event horizon corresponds to a
thermal bath, and AdS asymptotics to a conformal field theory.  The
AdS$_2$ infrared asymptotics of the Reissner-Nordstr\"om metric, 
which is supposed
to describe a finite-density, zero-temperature system, should
correspond to a (0+1)-dimensional conformal field theory, although the
nature of such a theory is not very clear.  It has been seen
explicitly in many calculations that the near-horizon geometry
influences the singular behavior of the inverse
propagators~\cite{Faulkner:2009wj}.

Nevertheless, the calculation of the full propagator always involves
the whole metric, not just its near-horizon
part~\cite{Faulkner:2009wj,Edalati:2010hk,Edalati:2010pn}.  One can
argue, rather generally, that the near-horizon geometry cannot
contain complete information about the long-distance physics.
Consider, for example, an extremal Reissner-Nordstr\"om black hole,
holographically dual to a finite-density medium.  This medium is
compressible, as seen from its equation of state, and should support a
gapless (for example, propagating or diffusive) mode related to charge
transport (these modes are seen explicitly in two-point Green
functions~\cite{Edalati:2010hk,Edalati:2010pn}).  However, the AdS$_2$
metric cannot support such a mode, as the spatial coordinates factor
out of it.  Another case is a holographic liquid where the infrared
metric is an AdS metric, but with a different speed of
light~\cite{Gubser:2008wz}.  In such a liquid the conformal Ward
identity $T^\mu_\mu=0$ (with the vacuum speed of light) should remain
valid in the long-distance regime.  But the near-horizon metric in
this case has a different speed of light than the one appearing in the
Ward identity, and it is not clear how  low-energy physics
``knows'' about the real light speed.

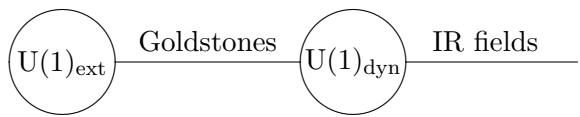
\begin{figure}[ht]
\begin{center}
\begin{picture}(215,40)(0,10)
\put(20,20){\circle{40}}
\put(20,14){\makebox(0,0)[b]{U(1)$_{\rm ext}$}}
\put(40,20){\line(1,0){70}}
\put(75,24){\makebox(0,0)[b]{Goldstones}}
\put(130,20){\circle{40}}
\put(130,14){\makebox(0,0)[b]{U(1)$_{\rm dyn}$}}
\put(150,20){\line(1,0){65}}
\put(180,24){\makebox(0,0)[b]{IR fields}}
\end{picture}
\end{center}
\caption{The moose diagram for holographic liquids.}
\label{fig:moose}
\end{figure}

In this paper, we suggest that the long-distance description of
holographic liquids involve a set of Goldstone bosons in addition to
the degrees of freedom living in the near horizon region.  We can
visualize the process of finding the low-energy effective theory as a
Wilsonian renormalization group procedure.  In this language, the
Goldstone boson appears as the only mode living outside the
near-horizon part of the metric that survives this procedure.  In the
simplest case of particle number diffusion, the Goldstone boson arises
from the spontaneous breaking of a U(1)$\times$U(1) symmetry down to
the diagonal U(1).  One of the U(1) is that of a conserved charge, but
the other U(1) is an emergent dynamical U(1) gauge field.  The
holographic infrared degrees of freedom, living in the near
horizon part of the metric (which, for shortness, will be called just
the IR degrees of freedom) are coupled to the dynamic U(1) field, but
are not coupled directly to the particle number U(1) field.  This is
summarized in the ``moose diagram'' of Fig.~\ref{fig:moose}.

Our picture is similar to the ``semiholographic'' models considered in
Ref.~\cite{Faulkner:2010tq}.  The difference is that
Ref.~\cite{Faulkner:2010tq} concerns mostly with probe fermion
fields, but we are interested in the degrees of freedom transporting
charge and energy-momentum.  The emphasis on the Goldstone modes
distinguishes this paper from other works seeking to relate the
properties of the boundary theory and the
horizon~\cite{Kovtun:2003wp,Iqbal:2008by,Bredberg:2010ky},

The dynamic gauge field connecting the Goldstone boson and the IR
fields bears some resemblance to the emergent gauge fields in some
condensed-matter models~\cite{deconfined}.  It suggests that the
holographic constructions and condensed matter models involving
emergent gauge fields are closer to each other than previously
thought.  Similar connections have been explored in
Ref.~\cite{Sachdev:2010um}.  One interesting fact that we found is the
appearance of dynamic gravity in the low-energy effective theories
arising from holography.  Note that there have been attempts to
construct lattice models that would give rise to gravity~\cite{Wen}.

The paper is organized as follows.  In Sec.~\ref{sec:gauge} we
consider the simplest problem: a gauge field in a fixed black-brane
metric.  We show that the diffusion mode can be interpreted as a
Goldstone boson, which is coupled, through an emergent gauge field, to
a stretched horizon with a finite electrical conductivity.  In
Sec.~\ref{sec:gravity} we tackle a more difficult problem of
gravitational fluctuations.  We show that the low-energy dynamics is
that of a Goldstone boson coupled to an emergent metric.  We show how
the viscosity of the stretched horizon becomes, through the Goldstone
boson, the viscosity at the boundary.  A by-product of this Section is
a bi-gravity formulation of hydrodynamics.  In
Sec.~\ref{sec:zero-sound} we give the Goldstone-boson interpretation
to the zero-temperature sound (zero sound) found in
Ref.~\cite{Karch:2008fa}.  We conclude with Sec.~\ref{sec:conclusion}.

\section{Diffusion from Goldstone boson dynamics}
\label{sec:gauge}

We illustrate the picture advocated above on the example of charge
diffusion at finite temperature.  The gravitational description
involves a gauge field $A_\mu$ in a black hole horizon,
\begin{equation}
  S = -\frac1{4g_{\rm YM}^2} \int\! d^5x\, \sqrt{-g}\, g^{\mu\alpha}
      g^{\nu\beta} F_{\mu\nu} F_{\alpha\beta}\,.
\end{equation}
The metric will be chosen in the form
\begin{equation}
  ds^2 = -r^2 f(r)dt^2 + r^2 d{\vec x}^2 + \frac{dr^2}{r^2f(r)}\,,
\end{equation}
where $f(r_0)=0$ at the horizon $r=r_0$ , and $f(\infty)=1$.  

\begin{figure}
\includegraphics[width=0.4\textwidth]{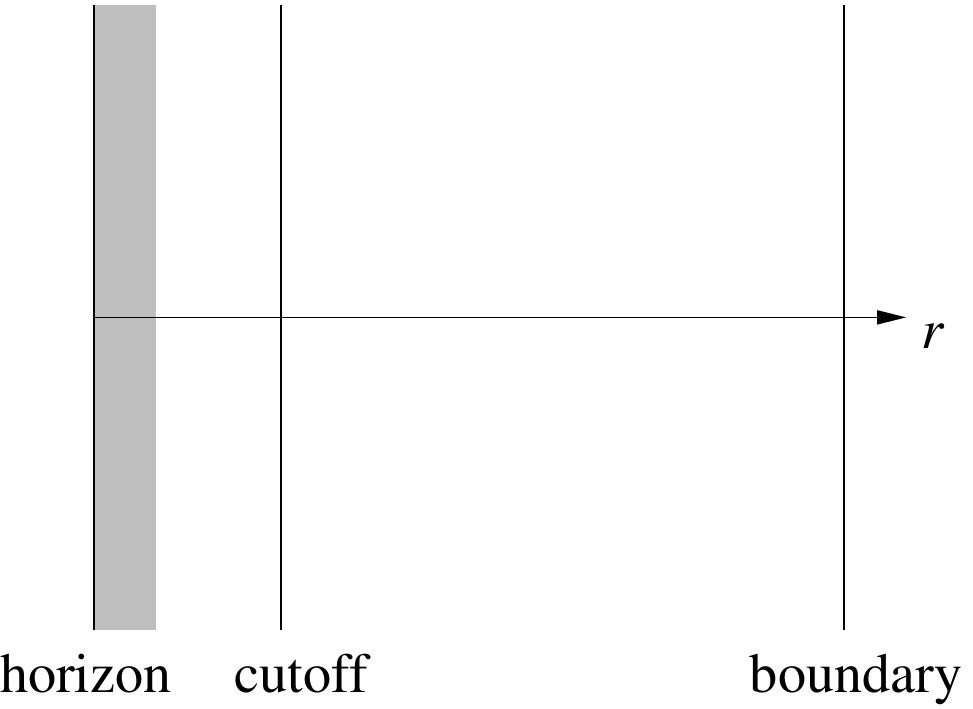}
\caption{The division of space into two regions.  The holographic IR modes 
live in the shaded region well below the cutoff.}
\label{fig:chopping}
\end{figure}

We are interested in the dynamics at distances larger than some scale.
This regime, following the holographic dictionary, maps onto a region
near the black hole horizon (the shaded region in
Fig.~\ref{fig:chopping}).  We then choose an arbitrary $r_\Lambda$ as
a coordinate separating the near-horizon region from the outside
region.  The action therefore is the sum of two actions,
\begin{equation}
  S = S_{\rm IR} + S_{\rm UV}.
\end{equation}
The gauge field $A_\mu=A_\mu(r=\infty)$ at the AdS boundary couples 
to $S_{\rm UV}$
only.  The value of $A_\mu$ at $r_\Lambda$, $a_\mu=A_\mu(r_\Lambda)$,
serves as a source for the IR theory.  This source also couples to the
UV degrees of freedom: the UV theory has two boundaries and is coupled
to two external gauge fields
\begin{equation}
  S = S_{\rm IR}[a_\mu, \phi_{\rm IR}] 
  + S_{\rm UV}[A_\mu,\,a_\mu,\phi_{\rm UV}].
\end{equation}
Here $\phi_{\rm UV}$ and $\phi_{\rm IR}$ denote fields in the UV and
IR theories, respectively.  The field $a_\mu$ should be determined by
the equation of motion, i.e., by the condition that the variation of
the total action with respect to it is zero: $\delta S/\delta
a_\mu=0$.  In the quantum theory, one should perform a path integral
over $a_\mu$.

The key observations are that (i) the UV theory is a confining theory
and can be rewritten as a theory of mesons, and (ii) the only meson
important at low energies is the Goldstone boson arising from the
breaking of U(1)$\times$U(1) symmetry down to the diagonal U(1) group.
This Goldstone boson arises in a manner similar to the pion in the
Sakai-Sugimoto model~\cite{Sakai:2004cn}.  To find the pions, 
we note that, if one fixes the values the temporal and spatial
components of the gauge field $A_\mu$ ($\mu\neq r$) on the two
boundaries, then the Wilson line
\begin{equation}
  \phi = \int_{r_\Lambda}^\infty\! dr\, A_r(r,x)
\end{equation}
is invariant under all gauge transformations preserving the boundary
values of $A_\mu$ up to a global transformation. The value of the
Wilson line is the Goldstone boson field.  Alternatively, if one works
in the radial gauge $A_r=0$, one
cannot impose the the Dirichlet boundary conditions $A_\mu=0$ ($\mu\ne
r$) on both boundaries.  If $A_\mu=0$ on one boundary, then it should
be $A_\mu=\d_\mu\phi$ on the other.  The Goldstone boson in the radial
gauge is that gauge parameter $\phi$.

Given the symmetries, we can write down the action
\begin{equation}\label{S-gauge-init}
  S = \int\!d^4x\,\frac12\!\left[f_t^2 (\d_0\phi-A_0+a_0)^2 
  - f_s^2 (\d_i\phi - A_i + a_i)^2\right]  + S_{\rm IR}[a_\mu].
\end{equation}
Here $f_t$ and $f_s$ are some low-energy constants that will be
determined later.
  
The IR theory $S_{\rm IR}[a_\mu]$ is defined, holographically, as the
theory dual to a U(1) field on a black brane horizon.  The only
information about this theory that we will need is its response to
$a_\mu$, which is an external gauge field from the point of view of
the IR degrees of freedom.  The relationship is found within the black
hole membrane
paradigm~\cite{Damour:1982,membrane-paradigm,Parikh:1997ma}, which
attaches a finite electrical conductivity $\sigma$ to the stretched
horizon at $r=r_\Lambda$,
\begin{equation}
  \mathbf{j}_{\rm IR} = \frac{\delta S_{\rm IR}}{\delta\mathbf{a}} =
  \sigma\mathbf{e}, \qquad e_i = -f_{0i} = -\d_0 a_i + \d_i a_0.
\end{equation}
In holography, the conductivity arises from the incoming-wave boundary
condition at the horizon~\cite{Kovtun:2003wp}.  We do not need to
specify the charge density $j^0=\delta S_{\rm IR}/\delta a_0$; it is
determined by the conservation law on the horizon, $\d_0j^0+\d_i
j^i=0$.

We pause here to clarify one subtlety. The dynamics on the horizon is
dissipative, therefore the equation $j=\delta S/\delta a$ is not valid
in the strict sense.  The precise meaning of this equation is found in
the closed-time-path formalism in the RA basis~\cite{Chou:1984es},
where $j$ is understood as $j_R$ and $a$ as $a_A$.  For the sake of
writing down the field equation, our naive equation is sufficient.

The extremization of the action Eq.~(\ref{S-gauge-init}) with respect
to $a_i$ gives (we set the external field $A_\mu=0$):
\begin{equation}
  \frac{\delta S}{\delta a_i} 
  = \frac{\delta S_{\rm UV}}{\delta a_i} +\frac{\delta S_{\rm IR}}{\delta a_i}
  \equiv j^i + j^i_{\rm IR}
  = - f_s^2(\d_i\phi + a_i) + \sigma e_i =0.
\end{equation}
This is one equation of motion.  The other equation of motion is
obtained by varying~(\ref{S-gauge-init}) with respect to $\phi$,
\begin{equation}
  f_t^2\d_0 (\d_0 \phi  + a_0) - f_s^2 \d_i (\d_i \phi  + a_i) =0. 
\end{equation}
We can write the equations in terms of the currents
$j^0=f_t^2(\d_0\phi+a_0)$, $j^i=-f_s^2(\d_i\phi+a_i)$:
\begin{align}
 & \d_0 j^0 + \d_i j^i = 0, \\
 & j^i = -\frac \sigma{f_s^2} \d_0 j^i - \frac\sigma{f_t^2} \d_i j^0.
   \label{ji-dj}
\end{align}
In the low-frequency regime ( $\omega\ll f_s^2/\sigma$), the first
term in the right-hand-side of Eq.~(\ref{ji-dj}) is negligible.  We
then obtain a diffusion equation for $j^0$,
\begin{equation}
  \d_0 j^0 - D\nabla^2 j^0 = 0,
\end{equation}
where the diffusion constant $D$ is related to the membrane electric
conductivity $\sigma$ and the susceptibility $f_t^2$ as
$D=\sigma/f_t^2$.  Note that $f_s^2$ does not enter this final
expression.

\subsection*{Calculating the parameters of the effective theory}

To find the value of $f_t^2$ and $f_s^2$, we match the effective field
theory with holographic calculations.  If we freeze the Goldstone
boson to $\phi=0$ and turn on constant external $A_0$ and $A_i$, then
the coefficients $f_t^2$ and $f_s^2$ are obtained by expanding $S$ to
quadratic order in the external fields,
\begin{equation}
  S = \frac12( f_t^2A_t^2 - f_s^2 A_i^2) .
\end{equation}
Freezing the Goldstone boson at $\phi=0$ corresponds to working in the
radial gauge $A_r=0$ and putting $A_\mu=0$ at the horizon.  The
equation satisfied by $A_t$ and $A_i$ are then
\begin{equation}
  \d_r (r^3\d_r A_t ) = 0, \qquad \d_r [r^3f(r)\d_r A_i] = 0.
\end{equation}
Solving the equations and substituting into the action, we then find
$f_t^2$ and $f_s^2$,
\begin{equation}\label{ftfs-int}
  f_t^2 = \frac1{g_{\rm YM}^2} \left[
      \int_{r_\Lambda}^\infty\!\frac{dr}{r^3} \right]^{-1},\qquad
  f_s^2 = \frac1{g_{\rm YM}^2} \left[
       \int_{r_\Lambda}^\infty\!\frac{dr}{r^3f(r)}\right]^{-1}.
\end{equation}

We notice here that $f_t^2$ remains finite in the limit $r_\Lambda\to
r_0$ but, since $f(r)$ vanishes linearly when $r\to r_0$, $f_s^2$
tends to zero logarithmically as $r_\Lambda\to r_0$.  Therefore, we
have to keep $r_\Lambda$ slightly outside the horizon radius $r_0$ in
our calculations.  In other words, we have to take the low-energy
(hydrodynamic) limit before the $r_\Lambda\to r_0$ limit.  The precise
value of $f_s^2$, however, is not important for the final value of the
diffusion constant.

\begin{comment}
\subsection{Unitary gauge and an alternative slicing of the action}

The connection to the diffusion equation can be made direct by going
to the unitary gauge $\phi=0$.  In this gauge, the whole discussion
can be abstracted away from the presence of the term
$f_s^2(\d_i\phi-a_i)^2$ in the Lagrangian.  In the unitarity gauge,
the conductivity equation on the horizon implies
\begin{equation}
  f_s^2 a_i = \sigma(\d_0 a_i - \d_i a_0)
\end{equation}
which implies that $\d_0 a_i\ll \d_i a_0$.  Therefore the current on
the horizon is only $\d_i a_0$.

To proceed, we use a slight different slicing of the action
\begin{equation}
  S = S_0 + \tilde S_{\rm IR}
\end{equation}
where $S_0$ now contains only the time derivative part of the
Goldstone action,
\begin{equation}\label{S0-gauge}
  S_0=\frac12f_t^2 (\d_0\phi -A_0+a_0)^2
\end{equation}
and $\tilde S_{\rm IR}$ combines the term in the Goldstone Lagrangian
with the IR action
\begin{equation}
  \tilde S_{\rm IR} [\phi, A_\mu, a_\mu] = -\frac{f_s^2}2 (\d_i\phi
    -A_i + a_i)^2 + S_{\rm IR}[a_\mu]
\end{equation}

For $\tilde S_{\rm IR}$, all we need to know is that
\begin{equation}
  \frac{\delta\tilde S_{\rm IR}}{\delta A_i} = \sigma(\d_0 a_i-\d_i
a_0) = \sigma \d_i a_0
\end{equation}
The conservation of current, in the unitary gauge, then becomes
\begin{equation}
  f_t^2 \d_0 a_0 - \sigma \nabla^2 a_0=0
\end{equation}
which has the form of a diffusion equation with the diffusion constant
$D=\sigma/f_t^2$.

\end{comment}

\section{Hydrodynamics and emergent gravity}
\label{sec:gravity}

We now generalize the discussion in Sec.~\ref{sec:gauge} to the case
of hydrodynamic modes in a finite-temperature plasma.  Instead of the
gauge field $A_\mu$ in the bulk, we now have the gravitational field.
The emergent U(1) gauge field $a_\mu$ is now replaced by gravitational
perturbations living on a surface near the horizon.  Hydrodynamics,
therefore, is a theory of a Goldstone boson, bifundamental with
respect to two gravities.  Such a Goldstone boson was considered in
Ref.~\cite{ArkaniHamed:2002sp}.  It is a map between the ``boundary
coordinates'' $x^\mu$ and ``horizon coordinates'' $X^M$.  Thus $X^M$
can be thought of as 4 scalar fields living on the boundary
coordinates $x^\mu$,
\begin{equation}
  X^M = X^M(x^\mu),
\end{equation}
and $x^\mu$ can be viewed as fields living on the horizon,
\begin{equation}
  x^\mu = x^\mu(X^M).
\end{equation}
We will use $\mu$ for the spacetime coordinates on the boundary, $M$
for spacetime coordinates on the horizon, $i,j$ for spatial
coordinates on the boundary and $a,b,\ldots$ for the spatial
coordinates on the horizon.  We assume the boundary to be a
four-dimensional spacetime, but the discussion can be generalized to
any number of dimensions.

The ground state corresponds to $X^M = \delta^M_\mu x^\mu$, around
which one can expand $X^M = \delta^M_\mu x^\mu + \phi^M(x^\mu)$.  The
fields $\phi^M$ then fluctuate around zero.

The Goldstone boson is coupled to the metric on the boundary
$g_{\mu\nu}$.  On the horizon, the metric is degenerate.  The $X^M$
space is what we will call a ``Galilei space,'' and is described in
the Appendix~\ref{sec:Gal-st}.  Such a space is characterized by a
degenerate metric $G_{MN}(X)$ and a null vector $n^M(X)$, so that
$G_{MN}n^N=0$.  Alternatively, one can describe the Galilei space in
terms of a spatial metric $G_{ab}$, a vector field $v^a$, and a
Galilei clock factor $\gamma$, which are all functions of $X^M$:
\begin{equation}
 ds^2 = G_{MN} dX^M dX^N = G_{ab}(dX^a -v^a dT) (dX^b -v^b dT), \qquad
 n^N = \frac1\gamma (1, v^a).
\end{equation}

\subsection{Ideal hydrodynamics as a theory of Goldstone bosons}

The action of the Goldstone boson should be invariant with respect to
reparametrization of $x^\mu$ and of $X^M$.  One could, in principle,
derive this action from the gravity action in the bulk.  We will,
however, guess the form of this action by improving on the previous
proposal of Ref.~\cite{Dubovsky:2005xd} (see also
Ref.~\cite{Leutwyler:1996er}).  We first introduce the notion of
$\det{}_{\!3}$.  Assume $A$ is a $4\times4$ matrix, then
\begin{equation}
  \det{}_{\!3}\, A =  \frac16(\tr A)^3-
  \frac12 \tr A \tr A^2 + \frac13 \tr A^3.
\end{equation}
The operation $\det{}_{\!3}$ is defined so that if $A$ is a matrix
with one zero eigenvalue, then $\det{}_{\!3}\,A$ is the product of
three other eigenvalues.  The action for the Goldstone boson is
\begin{equation}\label{S-Goldstone}
  S_0 = -\int\!d^4x\, \sqrt{-g}\, \epsilon\!
   \left(\sqrt{\det{}_{\!3}\, (OG)}\right),
\end{equation}
where $\epsilon(...)$ is a function of one variable, and the
$4\times4$ matrix $O$ is defined as
\begin{equation}\label{O-def}
  O^{MN} = g^{\mu\nu} \d_\mu X^M \d_\nu X^L,
\end{equation}
and ${(OG)^M}_N \equiv O^{ML}G_{LN}$.  Clearly, $S_0$ is invariant
with respect to diffeomorphisms of $x^\mu$ and $X^M$ spaces.

We can rewrite the action~(\ref{S-Goldstone}) in three-dimensional
language by introducing
\begin{equation}\label{ea-def}
  B^{ab}=g^{\mu\nu} e^a_\mu e^b_\nu, \qquad
  e^a_\mu = \d_\mu X^a - v^a \d_\mu T.
\end{equation}
A property of $B^{ab}$ is that $\tr_{4\times
4}(OG)^n=\tr_{3\times3}(BG)^n$, hence $\det_{\!3}(OG)=\det (BG)$.
Thus, the action can be written as
\begin{equation}\label{S0-detab}
  S_0 = -\int\!d^4x\, \sqrt{-g}\, \epsilon \Bigl( \sqrt{\det 
  B^{ab}} \sqrt{\det G_{ab}} \Bigr).
\end{equation}
If we set the metric to $G_{ab}=\delta_{ab}$, $v_a=0$, then the action
is the same as that of Ref.~\cite{Dubovsky:2005xd},
\begin{equation}\label{Dubovsky}
  S_0 = - \int\!d^4x\, \sqrt{-g}\, \epsilon \Bigl( 
  \det{}^{1/2} \bigl[g^{\mu\nu} \d_\mu X^a \d_\nu X^b\bigr]
  \Bigr).
\end{equation}
Equation~(\ref{S-Goldstone}) generalizes the
Lagrangian~(\ref{Dubovsky}) to take into account the coupling with the
emergent metric.  In Appendix~\ref{app:matching} we check, using
holography, that the Lagrangian~(\ref{S-Goldstone}) correctly encodes
the response of the fluid to homogeneous perturbations of the external
metrics.

The conventional formulation of relativistic fluid dynamics is
recovered in the unitary gauge $X^M=\delta^M_\mu x^\mu$.  In this
gauge, all the information about the fluid is contained in the horizon
metric $G_{MN}$.  It is clear from Eq.~(\ref{S0-detab}) that the
action depends only on four parameters: three components of $v^a$ and
$\det G_{ab}$.  In other words, $S_0$ has an additional gauge
invariance with respect to arbitrary changes of $G_{ab}$ that preserve
the determinant.  Using this extra invariance, we can fix the form of
the horizon metric to
\begin{equation}\label{Gs}
  G_{MN} = s^{2/3}(\eta_{MN} + u_M u_N),\qquad
  \eta_{MN}={\rm diag}(-1,1,1,1),
\end{equation}
where $u^M$ satisfies $(u^0)^2-(u^i)^2=1$.  The argument of $\epsilon$
in Eq.~(\ref{S-Goldstone}) then becomes $s$.  If we now identify
parameter $s$ with the local entropy density, $u^M$ with the local
fluid velocity, and $\epsilon(s)$ with the energy density (as a
function of the entropy density), then the stress-energy tensor,
computed by differentiating $S_0$ with respect to $g_{\mu\nu}$, has
the same form as the stress-energy tensor of a ideal fluid,
\begin{equation}
  T_0^{\mu\nu} = \frac{2}{\sqrt{-g}}\frac{\delta S}{\delta g_{\mu\nu}} =
  (\epsilon+P)u^\mu u^\nu + P\eta^{\mu\nu}.
\end{equation}
Here $P=s\epsilon'(s)-\epsilon$ is the pressure.  For a conformal
fluid, $\epsilon(s)\sim s^{4/3}$.

Note that Eq.~(\ref{Gs}) implies that the total entropy is equal to
the volume of the horizon, without the factor of $1/4G$, but the
factor can be reinserted without any problem.  We can also take the
alternative point of view that we set $4G=1$ in all formulas.

\subsection{Dissipation from coupling to Galilei metric}

The action~(\ref{S0-detab}) does not contain the coupling to shear
fluctuations of the horizon metric $G_{ab}$ (those which preserve
$\det G_{ab}$).  Thus, $S_0$ is analogous to the time-derivative term
$f_t^2(\d_0\phi-A_0+a_0)^2$ in the action~(\ref{S-gauge-init}) in
Sec.~\ref{sec:gauge}.  We therefore expect that it is not the full
Goldstone boson action $S_{\rm UV}$---there must be another term,
analogous to the spatial derivative term $f_s^2(\d_i\phi-A_i+a_i)^2$
in Eq.~(\ref{S-gauge-init}).  This term couples the Goldstone boson
with the shear fluctuations of $G_{ab}$, and will be call called
$S_{\rm shear}[X, g^{\mu\nu}, G_{MN}]$,
\begin{equation}
  S_{\rm UV} = S_0 + S_{\rm shear}\,.
\end{equation}
As in the gauge theory case where $f_s\to0$ as $r_\Lambda\to r_0$, we
expect $S_{\rm shear}$ to vanish in the limit $r_\Lambda\to r_0$, but
this limit cannot be taken before the hydrodynamic limit.

We will assume the most general form for $S_{\rm shear}$ dictated by
general coordinate invariance,
\begin{equation}
  S_{\rm shear} = \int\!d^4x\, \sqrt{-g}\,\, \L_1 \bigl(\tr (BG),\, 
  \tr (BG)^2,\, \tr (BG)^3 \bigr).
\end{equation}
Here $\L_1(x_1,x_2,x_3)$ is an arbitrary function of $x_n=\tr(BG)^n$.
We will limit ourselves to conformal field theories, so $\L_1$
transforms like $\L_1\to e^{4\omega}\L_1$ under Weyl transforms
$g_{\mu\nu}\to e^{-2\omega}g_{\mu\nu}$.  This means
\begin{equation}
  \sum_n n x_n \frac{\d \L_1}{\d x_n} = 2 \L_1.
\end{equation}
Since one can add an arbitrary constant to $\L_1$,  without losing
generality we can require that $\L_1=0$ when $BG$ is proportional to
the identity matrix, $BG=s^{2/3}\openone$,
\begin{equation}
  \L_1(x_n)|_{x_n=3s^{2n/3}}  = 0 .
%  \sum_n n \frac{\d \L_1}{\d x_n} &=0, 
\end{equation}
Conformal invariance then implies
\begin{equation}\label{ndL1n}
  \left. \sum_n n x_n \frac{\d\L_1}{\d x_n}\right|_{x_n=3s^{2n/3}} = 0.
\end{equation}

We shall assume that the new term $S_{\rm shear}$ favors energetically
configurations with equal eigenvalues of $BG$.  Thus in equilibrium
$BG=s^{2/3}\openone$; in the hydrodynamic regime the deviation from
equilibrium is small,
\begin{equation}\label{dOG}
  G = s^{2/3} B^{-1} + \delta G, \qquad \delta G \ll s^{-2/3} B^{-1}.
\end{equation}

Taking variations of $\L_1$ with respect to $g_{\mu\nu}$ and $G_{ab}$,
we find its contribution to the boundary and horizon stress-energy
tensors,
\begin{equation}\label{Tmn-OG}
  T_{\mu\nu}^{\rm shear} = g_{\mu\alpha} g_{\nu\beta}\frac{2}{\sqrt{-g}}
  \frac{\delta S}{\delta g_{\alpha\beta}}=
  - 2e^a_\mu e^b_\nu \sum_n n  
  \frac{\d\L_1}{\d x_n} [G(BG)^{n-1}]_{ab} + g_{\mu\nu} \L_1,
\end{equation}
where $e^a_\mu$ is defined in Eq.~(\ref{ea-def}), and
\begin{equation}\label{tab-OG}
  \tau_{ab}^{\rm shear} 
  = G_{ac} G_{bd} \frac{2}{\sqrt{-g}}
    \frac{\delta S_{\rm shear}}{\delta G_{cd}}=
  2\sum_n n \frac{\d\L_1}{\d x_n} [G(BG)^n]_{ab}.
\end{equation}
If we replace in Eqs.~(\ref{Tmn-OG}) and (\ref{tab-OG}) 
$G\to G_0\equiv s^{2/3}B^{-1}$,
then both stress tensors vanish, due to Eq.~(\ref{ndL1n}).
Therefore, both stress tensors are proportional to $\delta G$.
To relate them to each other, we notice that, to leading order in 
$\delta G$,
\begin{equation}
  \sum_n n \frac{\d\L_1}{\d x_n} G(BG)^n =
  \sum_n n \frac{\d\L_1}{\d x_n} G(BG)^{n-1} (BG_0)
  + \sum_n n \frac{\d\L_1}{\d x_n} G_0(BG_0)^{n-1} (B\delta G).
\end{equation}
The second sum in the right hand side vanishes due to Eq.~(\ref{ndL1n}).
We thus have
\begin{equation}
  \sum_n n \frac{\d\L_1}{\d x_n} G(BG)^n = 
  s^{2/3} \sum_n n \frac{\d\L_1}{\d x_n} G(BG)^{n-1},
\end{equation}
and the relationship between the boundary and the horizon
stress tensors arising from $S_{\rm shear}$ is
\begin{equation}\label{T1tau1}
  T^{\rm shear}_{\mu\nu} = - \Bigl(e^a_\mu e^b_\nu
  -\frac14 g_{\mu\nu}O^{ab}\Bigr) s^{-2/3}
  \tau^{\rm shear}_{ab} .
\end{equation}
Note that $T^{\rm shear}_{\mu\nu}$ is traceless.

We now relate the stress tensor $\tau_{ab}^{\rm shear}$ to the stress
tensor of the IR theory of the horizon degrees of freedom.  We make
the shear fluctuations of $G_{ab}$ dynamical, which means that the
variation of the action with respect to these fluctuations vanishes,
\begin{equation}
  \delta (S_{\rm shear} + S_{\rm IR}) = \int\!dx\,\sqrt{-g}\, 
  \tau_{\rm shear}^{ab}\delta G_{ab} 
  + \int\!dT\,d^3X\, \gamma\sqrt{G}\,\hat \tau^{ab}_{\rm hor}\delta G_{ab}=0,
\end{equation}
for all $\delta G_{ab}$ which satisfies $G^{ab}\delta G_{ab}=0$.  Here
$\hat\tau^{ab}_{\rm hor}$ is the stress tensor of the degrees of freedom
living on the horizon.  This implies
\begin{equation}\label{tau1tau}
  \tau_{\rm shear}^{ab} = 
  - \frac{\det |\d_\mu X^M| \gamma \sqrt{\det G_{ab}}}{\sqrt{-g}}
  (\hat\tau^{ab}_{\rm hor} + \lambda' G^{ab}),
\end{equation}
where $\lambda'$ is undetermined.  Moreover, the membrane
paradigm~\cite{Damour:1982,membrane-paradigm,Parikh:1997ma} implies
that the stress tensor at the horizon $\tau_{ab}$ is proportional to
the projected tensor $C_{ab}$ defined in the Appendix,
\begin{equation}\label{tauS}
  \hat\tau_{ab}^{\rm hor} = -\eta_0 \Bigl(C_{ab}- \frac13 G_{ab}C\Bigr) 
  - \zeta_0 G_{ab} C, \qquad 
  C_{ab} = \frac1\gamma(\nabla_a v_b + \nabla_b v_a + \dot G_{ab}),
  \quad C \equiv G^{ab} C_{ab},
\end{equation}
where $\eta_0$ and $\zeta_0$ are the shear and bulk viscosities of the
horizon.  From the ``membrane paradigm'' we have
$\eta_0=1/4\pi$ ($=1/16\pi G$), the value of $\zeta_0$ is not important
since the coefficient $\lambda$ in Eq.~(\ref{tau1tau}) is
undetermined.  Combining Eqs.~(\ref{T1tau1}), (\ref{tau1tau}), and
(\ref{tauS}), we find the additional contribution to the
stress-energy boundary tensor to be
\begin{equation}
  T_{\mu\nu}^{\rm shear} =  -\frac{\eta_0}{s^{2/3}}\, 
  \frac{\det |\d_\mu X^M| \gamma \sqrt{\det G_{ab}}}{\sqrt{-g}}
  \Bigl(e^a_\mu e^b_\nu - \frac14 g_{\mu\nu} O^{ab}\Bigr) 
  (C_{ab} + \lambda G_{ab}),
\end{equation}
where $\lambda$ is an undetermined coefficient.  This equation can be
rewritten in the 4-dimensional horizon form as
\begin{equation}\label{Tdiss}
  T_{\mu\nu}^{\rm shear} = -\frac{\eta_0}{s^{2/3}}\,
  \frac{\det |\d_\mu X^M| \gamma \sqrt{\det G_{ab}}}{\sqrt{-g}}\, 
   \Bigl(\d_\mu X^M \d_\nu X^N -\frac14 g_{\mu\nu} O^{MN}\Bigr) \, 
  (C_{MN} + \lambda G_{MN}).
\end{equation}

Now going to the unitary gauge $X^M=x^\mu\delta^M_\mu$, substituting
$G_{MN}=s^{2/3}(g_{MN}+u_M u_N)$, and choosing $n^M=u^M$ (using the fact
that $\lambda C_{MN}$ is independent of the clock factor $\gamma$),
we will find
\begin{multline}\label{Tmunu-l}
  T^{\mu\nu} = -\eta P^{\mu\alpha} P^{\nu\beta} 
  \Bigl(\d_\alpha u_\beta + \d_\beta u_\alpha 
    - \frac23 g_{\alpha\beta}\d\!\cdot\! u\Bigr)\\ 
  - \frac\eta{12} (g^{\mu\nu}+4u^\mu u^\nu) 
    P^{\alpha\beta}(\d_\alpha u_\beta + \d_\beta u_\alpha +
    2(u\!\cdot\!\d)\ln s+ \lambda P_{\alpha\beta}),
  \qquad \eta = \eta_0s,
\end{multline}
where $P^{\alpha\beta}=g^{\alpha\beta}+u^\alpha u^\beta$.  The second
term in the Eq.~(\ref{Tmunu-l}), proportional to $g^{\mu\nu}+4u^\mu
u^\nu$, can be absorbed into the ideal part of the stress-energy
tensor $(\epsilon+P)u^\mu u^\nu +Pu^{\mu\nu}=(g^{\mu\nu}+4u^\mu
u^\nu)P$ by a redefinition of the temperature. We reproduce here the
standard dissipative part of the stress-energy tensor, with the
viscosity equal to $\eta_0 s$.  Since on the horizon $\eta_0=1/4\pi$,
this implies $\eta/s=1/4\pi$.

\section{Holographic zero sound}
\label{sec:zero-sound}

In this section, we apply the philosophy developed above to a
zero-temperature case: the D3-D7 system at finite baryon density.  The
field theoretical description of such a system is the ${\cal N}=4$ super
Yang-Mills theory with ${\cal N}=2$ fundamental matter.  We assume the
number of matter flavors $N_f$ to be much smaller than the number of colors
$N_c$, $N_f\ll N_c$, so that the probe approximation works on the gravity
side of the duality.

The calculation of the current-current correlation function in this
system reveals a zero-temperature mode, which was called the
holographic zero-temperature sound, or, in analogy with a collective
mode in the Fermi liquid, the zero sound~\cite{Karch:2008fa}.  This
mode is different from zero-temperature collective excitations
encountered in many-body theory.  It has a linear dispersion relation
$\omega=vk$, with the velocity $v=(\d P/\d
\epsilon)^{1/2}$~\cite{Karch:2008fa,Kulaxizi:2008kv}.  Such a velocity
would be natural if the mode was the superfluid Goldstone boson
arising from the spontaneous breaking of the baryon number
symmetry~\cite{Son:2002zn}, but there is no indication that this
breaking takes place in the geometry.  Moreover, the damping rate of
the mode is $\Gamma\sim k^2$, with a coefficient which is not
suppressed by $N_c$, which is also inconsistent with the superfluid
Goldstone boson interpretation.

In light of what we know by now, the nature of the holographic zero sound
should be clear.  This mode is the Goldstone boson, but not of
the breaking of the global U(1) baryon symmetry, but rather one of the
breaking of a U(1)$_{\rm global}\times$U(1)$_{\rm gauge}$ symmetry
down to a diagonal U(1).  The imaginary part in the dispersion curve
of the Goldstone boson is due to the coupling of the dynamical U(1)
field to an infrared sector.

Our starting point will be the quadratic action for longitudinal
gauge-field fluctuations in the bulk (cf.\ Ref.~\cite{Karch:2008fa}):
\begin{equation}\label{S2}
  S = \frac{{\cal N}_q}2\! \int\! d^{p+1}x\, dz\, z^{2-p}
  \left[ f^3(z) (\d_z a_0 {-} \d_0 a_z)^2 + f(z) (\d_0 a_i {-}\d_i a_0)^2
   - f(z) (\d_z a_i {-} \d_i a_z)^2\right].
\end{equation}
Here $p$ is the number of spatial dimensions of the field theory, and
$f(z) = (1+z^{2p})^{1/2}$ (which corresponds to a fixed charge
density, equal to ${\cal N}_q$ times a dimensionless constant).  The
UV corresponds to $z=0$ and the IR to $z=\infty$.  Here we assume the
fundamental quarks to be massless.

We are interested in low-energy physics only.  We will choose some
value $z_\Lambda\gg1$, so that $1/z_\Lambda$ is the cutoff of the
low-energy effective theory.  The degrees of freedom of the theory can
be broken into the IR degrees of freedom, denoted collectively as
$\psi$, living in $z>z_\Lambda$; the ``emergent gauge field'' living
on the slice $z=z_\Lambda$, $a_\mu =a_\mu(z_\Lambda)$, and the UV
degrees of freedom, collectively denoted as $\phi_{\rm UV}$ living in
$z<z_\Lambda$.  The IR fields couple to $a_\mu$, while the UV fields
couple to both $A_\mu= a_\mu(0)$ and $a_\mu$.  The partition function of the
theory, in external fields, can be written as:
\begin{equation}
  Z[A_\mu] = \int\!D\psi\, Da_\mu\, D \phi_{\rm UV}\,
  \exp\left( iS_{\rm IR}[\psi, a_\mu] 
       + i S_{\rm UV}[\phi_{\rm UV}, a_\mu, A_\mu]\right).
\end{equation}

As discussed in Sec.~\ref{sec:gauge}, due to the IR cutoff at
$z_\Lambda$, $S_{\rm UV}$ describes an infinite tower of hadrons in a
confining theory.  In complete analogy with Sec.~\ref{sec:gauge}, the
only hadron relevant for the low-energy physics is the Goldstone boson
of the spontaneous breaking U(1)$\times$U(1)$\to$U(1).  Thus, the
effective action should now be
\begin{equation}\label{Seff}
  S = S_{\rm IR}[\psi, a_\mu] + \int\!d^4x\,\frac12\!\left[
      f_t^2(\d_0\phi -A_0+a_0)^2
      - f_s^2 (\d_i\phi-A_i+a_i)^2\right].
\end{equation}
The decay constants $f_t^2$ and $f_s^2$ can be determined by the same
method used to derive Eqs.~(\ref{ftfs-int}).  We find,
\begin{align}
  f_t^2 &= {\cal N}_q \left( \int_0^{z_\Lambda} \! \frac{dz}
           {z^{2-p}f^3(z)}\right)^{-1}
          ={\cal N}_q\frac{2\sqrt\pi\, p^2}{\Gamma\left( \frac1{2p}\right)
           \Gamma\left( \frac12-\frac1{2p}\right)}\,,\\
  f_s^2 &= {\cal N}_q \left( \int_0^{z_\Lambda} \! \frac{dz}
           {z^{2-p}f(z)}\right)^{-1}
          ={\cal N}_q\frac{2\sqrt\pi\, p} {\Gamma\left( \frac1{2p}\right)
           \Gamma\left( \frac12-\frac1{2p}\right)}\,.
\end{align}
(The integrals converge at large $z$, and the upper limit of
integration $z_\Lambda\gg1$ can be replaced by $\infty$.) 

Now we consider the IR sector.  We will first give the final
description of this sector, leaving the justification for later.  The
IR sector consist of an infinite set of (0+1)-d CFTs, 
one at each spatial point $\x$: $S_{\rm IR,
\mathbf{x}}$.  Each CFT therefore contains fields that depends only on 
time, which we denote collectively as $\psi_\x(t)$.
Each CFT contains operators $O_{\x,i}(t)$, $i=1\ldots p$ with
dimension 1, and also fields $\lambda_i$.  The whole Lagrangian is
\begin{align}\label{Seff2}
  & S = S_{\rm Goldstone} + {\cal N}_q\! \int\!d\x\!\int\!dt\,\left\{ 
  L_{\rm (0+1)dCFT}[\psi_\x(t)] + O_{i,\x}(t) \dot \lambda_{i,\x}(t) 
    + \lambda_{i,\x}(t) f_{0i}(t,\x)
   \right\},\\
  & S_{\rm Goldstone} = \int\!d^4x\,\frac12
    \left[ f_t^2 (\d_0\phi -A_0 + a_0)^2 - f_s^2
    (\d_i\phi -A_i + a_i)^2\right]\,,
\end{align}
where we factor out ${\cal N}_q$ from the Lagrangian of the (0+1)d CFT.

Integrating out the $\psi$ degrees of freedom, one gets the following
effective Lagrangian
\begin{equation}
  S = S_{\rm Goldstone} +
  {\cal N}_q\! \int\!d\x\, \Bigl[\frac i2 \int\!\frac{d\omega}{2\pi}\,
    |\omega|^3 |\lambda_{i,\x}(\omega)|^2 
  + \int\!dt\, \lambda_{i,\x}(t) f_{0i}(t,\x)\Bigr],
%  + \int\!d^4x\,\frac12\left[ f_t^2 (\d_0\phi - a_0)^2 - f_s^2
%    (\d_i\phi - a_i)^2\right],
\end{equation}
and integrating over $\lambda_i$, one gets
\begin{equation}
  S= S_{\rm Goldstone} + \frac {{\cal N}_q}2\! \int\!\frac{d^4q}{(2\pi)^4}\,
   \frac i{|\omega|^3}|f_{0i}(\omega,{\bf q})|^2.
%  + \frac{f_t^2}2 (\d_0\phi - a_0)^2 + \frac{f_s^2}2
%    (\d_i\phi - a_i)^2
\end{equation}
Now choosing, e.g., the $a_0=0$ gauge, and diagonalizing a $2\times2$
matrix for $\phi$ and $a_i$, one gets the dispersion relation for the
zero sound,
\begin{equation}
  \omega = v q - i\gamma q^2, \qquad
  \gamma = \frac{f_s^2 v^2}{2{\cal N}_q}\,,
\end{equation}
where $v=f_s/f_t=1/\sqrt p$, and the coefficient
$\gamma$ in the imaginary part is
\begin{equation}
  \gamma = \frac{f_s^2v^2}{2{\cal N}_q} = \frac{\sqrt\pi}
  {\Gamma\left( \frac1{2p}\right)
           \Gamma\left( \frac12-\frac1{2p}\right)}\,.
\end{equation}
The result for the dispersion relation coincides with that of
Ref.~\cite{Karch:2008fa}, which is a check of the validity of the
effective theory~(\ref{Seff2}).

Now let us justify Eq.~(\ref{Seff2}).  In the $a_z=0$ gauge, the field
equations in the full metric are
\begin{align}
   (f^3 z^{2-p}a_0')' -fz^{2-p}(q^2a_0 +\omega q_i a_i) &= 0,\\
   (f z^{2-p}a_i')' + fz^{2-p}(\omega q_i a_0 + \omega^2 a_i) &= 0.
\end{align}
In the regime $z\gg1$, $f=z^p$.  One can see that $a_0$ changes with
$z$ so slowly that it can be considered a constant.  Then
$\tilde a_i=a_i +q_ia_0/\omega$ satisfied the equation
\begin{equation}
  (z^ 2 \tilde a_i')' + z^2 \omega^2 \tilde a_i = 0.
\end{equation}
Changing variable to $\phi_i=z \tilde a_i$, we see that $\phi$ satisfies
the equation of a massless scalar in AdS$_2$.  The action for $\phi_i$ is
\begin{equation}\label{Sphi}
  S = \frac{{\cal N}_q}2\! \int\!d^{p+1}x\, dz\, 
      [(\d_0\phi_i)^2 - (\d_z\phi_i)^2].
\end{equation}
There are two CFTs corresponding to
(\ref{Sphi})~\cite{Klebanov:1999tb}.  In the first CFT the operator
$O$ dual to $\phi$ has dimension 1 and correlation function (in
Euclidean space) $\< O O \> = {\cal N}_q |\omega|$; in the second CFT,
$O$ has dimension 0 and $\< O O \> = {\cal N}_q |\omega|^{-1}$.  (The
coupling of $O$ and $\phi$ is taken to be ${\cal N}_q\phi O$, so that
${\cal N}_q$ factors out of the action.)

To determine the dimension of the operator dual to $\phi$, let us
first assume $a_0=0$, for simplicity, so $\tilde a_i=a_i$.  The boundary
condition for $\phi$, for $1\ll z \ll 1/\omega$, is
$\phi=a_iz+\cdots$, which is the more regular asymptotics near the
boundary (the other one is $z^0$).  Therefore the emergent electric
gauge field $a_i$ serves as the source for the operator dual to
$\phi$, and the dimension of that operator is 0.  Hence our model
is
\begin{equation}
  S = {\cal N}_q \Bigl[ S_{\rm (0+1)d CFT} 
      -\int\!d\x\!\int\!dt\, a_i(t,\x) O_{i,\x}^{\Delta=0}(t) \Bigr]
      + S_{\rm Goldstone}.
\end{equation}
This is the action written in the $a_0=0$ gauge.  To restore gauge
invariance, we can introduce a Legendre multiplier to enforce the
constraint $\d_t \tilde a_i = f_{0i}$:
\begin{equation}
  S = {\cal N}_q \Bigl[ S_{\rm (0+1)d CFT} 
  - \int\!d\x\!\int\!dt\, \tilde a_i(t,\x) O_{i,\x}^{\Delta=0}(t)
      - \!\int\!d^4x\,\lambda_{i,\x} (\d_t \tilde a_i(x)-f_{0i}(x))\Bigr] 
   + S_{\rm Goldstone}.
%      + \frac{f_t^2}2 (\d_0 \phi -a_0)^2 
%      - \frac{f_s^2}2 (\d_i \phi -a_i)^2
\end{equation}
Now we note that to integrate over $\tilde a_i$ is to take a Legendre
transform and convert $(0+1)d$ CFT into a CFT with scalar operator of
dimension 1.  In this way we arrive to Eq.~(\ref{Seff2}).

\begin{comment}

\section{A low-energy effective theory for the holographic quantum 
liquid dual to the Reissner-Nordstr\"om back hole}

In this final section, we speculate about the nature of the
holographic liquid dual to the Reissner-Nordstr\"om black hole.

We expect that the UV part of the Lagrangian in this case is still the
fluid Lagrangian.  The U(1) gauge field, to leading order, is coupled
to the entropy density,
\begin{equation}
  S = \int\!d^4x\, \sqrt{-g}\, f ( O ) 
  + (\d_\mu\phi -A_\mu) \d_\nu X^a \d_\lambda X^b \d_\rho X^c
\end{equation}

\end{comment}

\section{Conclusion}
\label{sec:conclusion}

In this paper, we have been advocating the point of view that
holographic liquids can be described, at long distances, by a theory
of Goldstone bosons coupled to an infrared sector through emergent
gauge and gravitational fields.  We consider in this paper only a few
simplest examples.  It should be possible to extend the calculation in
this paper to other cases, for example for the R-charged black holes,
where the relationship between boundary and horizon kinetic
coefficients is not trivial~\cite{Jain:2010ip}.  Possibly, the most
interesting applications of our formalism are zero-temperature
systems: the holographic superfluids and the system dual to the
extremal Reissner-Nordstr\"om black hole.  The latter plays a central
role in recent construction of holographic non-Fermi liquids.  In the
case of the extremal Reissner-Nordstr\"om black holes, it has been
found that the Kubo's formulas yield finite values for the kinetic
coefficients (for example, the shear viscosity
$\eta$)~\cite{Edalati:2009bi}.
%and the two-point functions of the
%stress-energy tensor have the same form as in hydrodynamics~\cite{}.  
However, the effective low-energy description of extremal
Reissner-Nordstr\"om black holes cannot be hydrodynamics.  In
hydrodynamics, there is a formula for entropy production (in local
fluid rest frame)
\begin{equation}
  \d_\mu s^\mu = \frac \eta T 
  \Bigl(\d_i u_j + \d_j u_i - \frac23 \d\cdot u \Bigr)^2.
\end{equation}
This formula does not make sense if $\eta$ is finite in the limit
$T\to 0$: the rate of entropy production would be infinite.  The
effective field theory therefore has to be of a different nature.  It
seems that the effective theory has to involve Goldstone modes,
coupled with AdS$_2$ degrees of freedom.  However, the details of this
theory need to be worked out.

The new point of view on holographic liquids reduces the problem of
finding the low-energy dynamics of such liquids into finding the
Goldstone boson degrees of freedom, the horizon degrees of freedom,
and the manner they are coupled together.  The appearance of the
emergent gauge fields brings an interesting questions about the
possible relationships between recent constructions of holographic
liquids with the older attempts to construct nontrivial low-energy
effective theories of strongly correlated electrons or spin systems,
which typically involve a ``deconfinement'' of emergent gauge fields.
Hopefully, our work will help bridging the gap between holographic
models and the field-theoretical models for strongly coupled
electronic systems.

The authors thank A.~O'Bannon, A.~Karch, Hong Liu, J.~Polchinski, and
A.~Strominger for discussions.  This work is supported, in part, by
DOE grant DE-FG02-00ER41132.

\appendix

\section{Galilei spacetime and Galilei field theories}
\label{sec:Gal-st}

\subsection{Galilei spacetime}

By ``Galilei spacetime'' we have in mind a structure consisting of
manifold with a degenerate metric
\begin{equation}
  ds^2 = G_{MN} dx^M dx^N, \qquad G_{MN} n^M = 0,
\end{equation}
and a Galilean clock factor $\gamma(T,X)$.  We will say that the
combination $(G_{MN},\gamma)$ defines a Galilei spacetime.  The null
metric can be parameterized by the null vector $v^a$ and a spatial
metric $G_{ab}$,
\begin{equation}\label{ds2-Gv}
  ds^2 = G_{ab} (dX^a - v^a dT)(dX^b - v^b dT),
\end{equation}
and so the Galilei space can be said to be characterized by $(G_{ab},
v^a, \gamma)$.

The Galilei spacetime can be considered as a limit $\epsilon\to0$ of a
spacetime with a metric
\begin{equation}\label{ds2e}
  ds^2_\epsilon = G_{MN}^\epsilon dX^M dX^N = 
  -\epsilon^2\gamma^2 dT^2 + G_{ab} (dX^a - v^a dT)(dX^b - v^b dT).
\end{equation}
All quantities for the Galilei structure should be defined to be
finite in the limit $\epsilon\to0$.  For example, the volume element
is defined as
\begin{equation}
  \lim_{\epsilon\to0} \frac1\epsilon\sqrt{-\det G^{\epsilon}_{MN}}\, d^4X
  = \gamma \sqrt{G} \, dT\,d^3X,
\end{equation}
where $G\equiv\det G_{ab}$.  The general coordinate transformations
(diffeomorphisms) of the Galilei spacetime can be obtained as the
$\epsilon\to0$ limit of the diffeomorphisms on the space~(\ref{ds2e}).
One can easily work out the action of infinitesimal diffeomorphisms
on the metric components.  Under spatial transforms, $X^a\to
X^{a\prime}=X^a+\xi^a$,
\begin{align}
  \delta G_{ab} &= -\xi^c \d_c G_{ab} - G_{cb} \d_a \xi^c - 
     G_{ac} \d_b \xi^c,\\
  \delta v^a &= -\xi^c \d_c v^a + v^c \d_c \xi^a + \dot \xi^a,\\
  \delta \gamma &= -\xi^c \d_c \gamma,
\end{align}
and under time transforms, $T\to T'= T+ \xi$,
\begin{align}
  \delta G_{ab} &= -\xi \dot G_{ab} + v_a \d_b \xi + v_b \d_a \xi, \\
  \delta v^a &= -\xi \dot v^a - v^a d_T \xi, \\
  \delta \gamma &= -\xi \dot \gamma - \gamma d_T \xi,
\end{align}
where $d_T\xi\equiv \dot\xi + v^c\d_c \xi$.  These can be taken as the
intrinsic definition of diffeomorphisms of the Galilei space, without
referring to the limiting procedure $\epsilon\to0$.  The action of the
Goldstone boson should be invariant with respect to diffeomorphisms of
both the physical spacetime $x^\mu$ and the Galilei spacetime $X^M$.

Under diffeomorphisms, contravariant vectors and tensors transform as
\begin{align}
  \delta A^M & = -\xi^L \d_L A^M + A^L \d_L \xi^M, \\
  \delta A^{MN} &= -\xi^L \d_L A^{MN} + A^{LN} \d_L \xi^M
                   + A^{ML} \d_L \xi^N,
\end{align}
while covariant vectors and tensors transform as
\begin{align}
  \delta A_M & = -\xi^L \d_L A^M - A_L \d_M \xi^L, \\
  \delta A_{MN} &= -\xi^L \d_L A^{MN} - A_{LN} \d_M \xi^L
                   - A_{ML} \d_N \xi^L.
\end{align}

The Galilei space possesses one intrinsic vector field
\begin{equation}
  n^M = \frac1\gamma (1, \, v^a).
\end{equation}
One can check that $n^M$ transforms like a vector under
diffeomorphisms.  It is a null vector: $G_{MN} n^M=0$.  In fact one
can take the pair $(G_{MN},n^M)$ as the definition of the Galilei
space.

Since the Galilei metric is degenerate, indices can be lowered using
$G_{MN}$ but, in general, cannot be raised.  Tensors obtained by
lowering the indices of a fully contravariant tensor are perpendicular
to the null vector,
\begin{equation}
  A_{MN} = G_{MK} G_{NL} A^{KL} \Rightarrow n^M A_{MN} = n^N A_{MN}=0.
\end{equation}
Such a tensor is completely determined by its spatial components:
$A_{0a} = -A_{ab} v^b$, $A_{00}= A_{ab} v^a v^b$.  One can regard the
spatial three-tensor $A_{ab}$ as an object by itself, which we will
call a projected tensor.  Under spatial reparametrization it
transforms as a conventional tensor in three-dimensional space, and
under time reparametrization it transforms as
\begin{equation}
%  X^c\to X^c+\xi^c: &\quad \delta \hat A_{ab} = -\xi^c \d_c \hat A_{ab} 
%     - \d_a \xi^c \hat A_{cb} - \d_b \xi^c \hat A_{ac}\\
  T\to T+\xi: \quad \delta \hat A_{ab} = -\xi \d_t \hat A_{ab} 
     + \d_a \xi v^c \hat A_{cb} + \d_b \xi v^c \hat A_{ac}.
\end{equation}
The metric tensor $G_{ab}$ is one such tensor.  The indices of a
projected tensor can be raised by using the inverse spatial metric
$G^{ab}$: $\hat A^{ab} = G^{ac} G^{bd} A_{cd}$.  This fully
contravariant projected tensor transforms under time reparametrization
as
\begin{align}
%  X^c\to X^c+\xi^c: &\quad \delta \hat A^{ab} = -\xi^c \d_c \hat A^{ab} 
%     + \d_c \xi^a \hat A_{cb} + \d_c \xi^b \hat A_{ac}\\
  T\to T+\xi: \quad \delta \hat A^{ab} = -\xi \d_t \hat A^{ab} 
     - v^a \d_c \xi \hat A^{cb} - v^b \d_c \xi \hat A^{ac}.
\end{align}
$G^{ab}$ is a contravariant projected tensor.  Note that a
contravariant projected tensor does not corresponds uniquely to a
four-tensor, rather, it corresponds to a whole class of four-tensors
which differ from each other by $A^{MN}\to A^{MN} + n^M k^N + n^N
k^M$.

We can construct, in analogy with the extrinsic curvature, the
following symmetric tensor,
\begin{equation}
  C_{MN} = 2 \nabla_{(M} n_{N)}  = 
  2 G_{L(M}\d_{N)} n^L + n^L \d_L G_{MN}.
\end{equation}
Since $n^M C_{MN}=0$, $C_{ab}$ is a projected tensor.  In components,
\begin{equation}\label{sigma-comp}
  C_{ab} = \frac 1\gamma (2 \nabla_{(a} v_{b)} + \dot G_{ab}).
\end{equation}
This tensor is proportional to the inverse of the Galilei
clock factor $\gamma$.

\subsection{Stress-energy tensor in Galilei field theories}

For a quantum field theory in Galilei space, one can define the
stress-energy tensor by taking small variation of the action with
respect to the external metric,
\begin{equation}\label{dS-MN}
  \delta S= \frac12\int\! dT\, d^3X\, \gamma \sqrt{\det G_{ab}} \,
         \tau^{MN} \delta G_{MN}.
\end{equation}
Since the matrix $G_{MN}$ is constrained to be degenerate, the
stress-energy tensor $\tau^{MN}$ is defined up to one arbitrary
contribution,
\begin{equation}
   \tau^{MN} \to \tau^{MN} + \lambda n^M n^N,
\end{equation}
so there are 9 independent components of $\tau^{MN}$ in four
dimensions.  By lowering the indices $\tau_{MN} = G_{MA} G_{NB}
\tau^{AB}$, one obtains a transverse tensor $\tau_{ab}$.  Note that
$\tau_{ab}$ contains less information than $\tau^{MN}$: there are
three extra independent components in $\tau^{MN}$.  This can be seen
by rewriting Eq.~(\ref{dS-MN}) in components,
\begin{equation}
  \delta S = \int\!dT\,d^3X\, \gamma\sqrt{\det G_{ab}} \, 
  \Bigl(\frac12 \hat \tau^{ab} \delta G_{ab} +\rho_a \delta v^a\Bigr), 
\end{equation}
where
\begin{align}
  \hat \tau^{ab} &= \tau^{ab} - v^a \tau^{0b} - v^b \tau^{a0} 
   + v^a v^b \tau^{00},\\
  \rho_a &=  v_a \tau^{00} - G_{ab} \tau^{0b}.
\end{align}

\section{Matching effective theory with holography}
\label{app:matching}

In order to compare and match our discussion to an actual AdS/CFT
calculation, we first work out the on-shell action of the holographic
setup to quadratic order in metric fluctuations with boundary
conditions imposed at the boundary as well as an intermediate cutoff
scale.  Following Refs.~\cite{Policastro:2002se,Policastro:2002tn} and
its conventions, the thermal AdS background is given by
\begin{equation}
  ds^2 = \frac{(\pi T R)^2}u 
  \left(-f(u)dt^2+d{\vec x}^2\right)+\frac{R^2}{4u^2f(u)}du^2
  = g^{(0)}_{MN}dX^MdX^N ,
\end{equation}
where $f(u)=1-u^2$.  The boundary is at $u=0$ and the horizon at
$u=1$.  We denote by $u_\Lambda$ the position of the stretched
horizon, which separates the UV and IR parts of the metric.  For the
fluctuations, defined through $g_{MN}=g^{(0)}_{MN}+h_{MN}$, we
introduce the parameterization
\begin{equation}
  H_{tt} = \frac{u h_{tt}}{f(\pi T R)^2}\,, \quad
  H'_{uu} = \frac{u\sqrt{f}h_{uu}}{R^2}\,, \quad
  (H_{ij}, H_{ti}, H_{u\mu}) = \frac u{(\pi T R)^2}
  (h_{ij}, h_{ti}, h_{u\mu}).
%  \quad
%  H_{ti} = \frac{u h_{tj}}{(\pi T R)^2}\,, \quad
%  H_{u\mu} = \frac{u h_{u\mu}}{(\pi T R)^2}\,.
\end{equation}
In the gauge $H_{uM}=0$ the boundary values of $H_{\mu\nu}$ in the
on-shell action are then the sources of the dual stress-energy tensor.

As an aside and simple observation: The length $
s=\int_{\tau_0}^{\tau_1}\!d\tau\sqrt{g_{MN}\d_\tau X^M\d_\tau X^N}$ of
a trajectory $X^\mu(\tau)=u(\tau) \delta^\mu_u$ is shifted to linear
order by $\delta s=L(H_{uu}(u(\tau_1))-H_{uu}(u(\tau_0)))$.

We want to turn on constant external metric perturbations, keeping the
Goldstone fields frozen at the vacuum value in
Eq.~(\ref{S-Goldstone}).  This is equivalent to fixing the boundary
conditions $H_{\mu\nu}(0)=h_{\mu\nu}$, $H_{\mu\nu}(u_\Lambda)=0$ in
the gauge $H_{u\mu}=0$ ($\mu,\nu\neq u$).  The component $H_{uu}$
requires a special treatment (see below).

Static, spatially homogeneous fluctuations decouple according to their
respective spin.  Spin-one fluctuations spanned by $H_{ti}$, $H_{ui}$
and spin-two fluctuations spanned by the components of
$\tilde{H}_{ij}=H_{ij}-\frac{1}{3}\delta_{ij}H_{kk}$ satisfy the
linearized equations of motions
\begin{equation}
\begin{split}
  0 &= H_{ti}''(u)-\frac{1}{u}H_{ti}'(u) , \\
  0 &= \tilde{H}_{ij}''(u)-\frac{1+u^2}{uf(u)}\tilde{H}_{zx}'(u) .
\end{split}
\end{equation}
The components $H_{ui}$ drop out to linear order and can be consistently
set to zero.  With the boundary conditions $H(0)=h$,
$H(u_\Lambda)=0$, the equations are solved by
\begin{equation}
\begin{split}
  H_{ti}(u) &= h_{ti} \left( 1- \frac{u^2}{u_\Lambda}\right) , \\
  \tilde H_{ij}(u) &=
  \tilde h_{ij} \left( 1- \frac{\ln f(u)}{\ln f(u_\Lambda)}\right) .
\end{split}
\end{equation}
Since static and homogeneous gauge transformations in these channels
are generated by Killing vectors of the background, gauge
transformations do not impose additional constraints on these solutions.

For the spin zero fluctuations, spanned by $H_{tt}$, $H_{ii}$, $H_{ut}$
and $H'_{uu}$, the linearized Einstein equations yield
\begin{equation}
\begin{split}
  0 &= Z''(u)-\frac{1}{u}Z'(u) ,\\
  0 &= (2+f(u))H_{ii}'-3f(u)H_{tt}'(u)+24\sqrt{f(u)}H'_{uu}(u) ,
\label{eq:AdSfluctuation_scalar}
\end{split}
\end{equation}
where we introduced the gauge-invariant combination
$Z(u)=(1+u^2)H_{ii}(u)+3f(u)H_{tt}(u)$. Similar as in the spin one
case, $H_{ut}$ drops out to linear order and can be set to zero.
However, since one equation in (\ref{eq:AdSfluctuation_scalar}) is
first order, we cannot set $H_{uu}(u)=0$.  For this reason we keep
$H'_{uu}(u)$ arbitrary for the moment and recall that each choice of
it defines a separate gauge.  The solution for the equation for $Z$ is
\begin{equation}
  Z(u) = Z(0)\left( 1 - \frac{u^2}{u_\Lambda^2}\right),
\end{equation}
and hence we have
\begin{equation}
\begin{split}
  0 &= (2+f(u))H_{ii}'(u)-3f(u)H_{tt}'(u)+24\sqrt{f(u)}H'_{uu}(u),\\
  0 &= (1+u^2)H_{ii}(u)+3f(u)H_{tt}(u) - Z(0) 
       \left(1 - \frac{u^2}{u_\Lambda^2}\right),
\end{split}
\end{equation}
which yield
\begin{equation}
\begin{split}
  H_{tt}(u) &= \frac{(1+u_\Lambda^2)}{6u_\Lambda^2}Z(0)
    +\frac{2(1+u^2)H_{uu}(u)}{\sqrt{f(u)}}\,,\\
  H_{ii}(u) &= \frac{(-1+u_\Lambda^2)}{2u_\Lambda^2}Z(0)
    - 6\sqrt{f(u)}H_{uu}(u) \,.
\end{split}
\end{equation}
To satisfy the boundary condition
$H_{tt}(u_\Lambda)=H_{ii}(u_\Lambda)=0$, we require
\begin{equation}
  H_{uu}(u_\Lambda) = - \frac{\sqrt{f(u_\Lambda)}}{12u_\Lambda^2} Z(0).
\end{equation}
Then the boundary conditions $H_{tt}(0)=h_{tt}$ and $H_{ii}(0)=h_{ii}$
can be achieved by choosing appropriate $Z(0)$ and $H_{uu}(0)$.  It is
worth noting that we can choose the metric perturbation and
derivatives in $u$ to vanish at both boundaries and that
$H_{uu}(0)-H_{uu}(u_\Lambda)$ shows up in the length of the trajectory
mentioned above.

The gravity action is given by the sum
\begin{equation}
  S = S_{\rm EH}+S_{\rm GH}+S_{\rm CT} \,,
\end{equation}
where the Einstein-Hilbert term $S_{\rm EH}$, the Gibbons-Hawking term
$S_{\rm GH}$ and the counter-term $S_{\rm CT}$ are defined as
\begin{equation}
\begin{split}
  S_{\rm EH} &= \phantom{-} \frac{N^2}{8\pi^2R^3}
  \int_{u_\Lambda}^{u_\epsilon}\!du\,d^4x\,\sqrt{-g}
   \left(\mathcal{R}+\frac{12}{R^2}\right) , \\
  S_{\rm GH} &= \phantom{-} \frac{N^2}{4\pi^2R^3}
  \int\!d^4x\,\left.\sqrt{-\gamma}K\right|_{u_\Lambda}^{u_\epsilon} ,\\
  S_{\rm CT} &= -\frac{3N^2}{4\pi^2R^4}
  \int\!d^4x\,\left.\sqrt{-\gamma}\right|^{u_\epsilon} .
\end{split}
\end{equation}
Here we introduced $u_\epsilon$ as a regulator of the renormalization
scheme in order to have finite intermediate results and will take
$u_\epsilon\to 0$ at the end of the calculation. Also note that the
counterterm only contributes for $u=u_\epsilon$. Since the
Einstein-Hilbert action also decomposes into surface terms for
fluctuations obeying the equations of motions, i.e.
$S_{\rm EH}=S^{\rm boundary}_{{\rm EH}, u=u_\epsilon} +
S^{\rm boundary}_{{\rm EH}, u=u_\Lambda}$,
we have two contributions to the on-shell action:
\begin{equation}
  S = \underbrace{ (S^{\rm boundary}_{{\rm EH}, u=u_\Lambda}
      + S_{\rm GH,u=u_\Lambda})}_{\equiv S_{\Lambda}}
    +\underbrace{ (S^{\rm boundary}_{{\rm EH}, u=u_\epsilon}
   +S_{\rm GH, u=u_\epsilon}+S_{\rm CT, u=u_\epsilon})}_{\equiv S_{\epsilon}}.
\end{equation}

The evaluation of the on-shell action is tedious and we only quote the
results.  Requiring $H_{uu}'(0)=H_{uu}''(0)=0$ we find for the
contribution from the boundary in the limit $u_\epsilon\to0$
\begin{multline}
  S_\epsilon = \frac{\pi^2N^2T^4 V}8 \Bigl[
    -1 + \frac 12( h_{ii} + 3h_{tt} )+\frac 1{24}\bigl(h_{ii}^2 - 36 h_{ti}^2
    - 6\tilde h_{ij} \tilde h_{ij} 
    + 6 h_{ii} h_{tt} + 9 h_{tt}^2 \\
    \phantom{=\frac{\pi^2N^2T^4}8\biggl(}
    - 8 h_{ii} H_{ii}''(0) + 12 h_{tt} H_{ii}''(0)
    - 24 h_{ti} H_{ti}''(0) 
%  \phantom{=\frac{\pi^2N^2T^4}8\biggl(}
  + 12 h_{ij} \tilde H_{ij}''(0) + 12 h_{ii} H_{tt}''(0)\bigr)\Bigr].
\end{multline}
Here we already used the requirement that $H_{\mu\nu}(u)$ is finite at
the boundary and even in $u$.  The contribution $S_\Lambda$ in the
limit $u_\Lambda\to1$ with $H_{uu}'(u_\Lambda)=H_{uu}''(u_\Lambda)=0$
and $H_{\mu\nu}=0$ vanish: $S_\Lambda=0$.  Therefore, after plugging
in the equations of motions we obtain
\begin{equation}
  S = \frac{\pi^2N^2T^4V}8 \Bigl[ -1 + \frac12 ( h_{ii} + 3 h_{tt})
   + \frac1{24}\bigl( -3 h_{ii}^2 + 12 h_{ti}^2 - 6 \tilde h_{ij}^2 
   - 6 h_{ii} h_{tt} + 9 h_{tt}^2 \bigr) \Bigr].
\end{equation}

This expression should be compared to the expansion of
Eq.~(\ref{S-Goldstone}) in unitary gauge using $\epsilon(s)= C
s^{4/3}$. For $X^M=\delta^M_\mu x^\mu$,
$g_{\mu\nu}=\eta_{\mu\nu}+h_{\mu\nu}$ and
$G_{MN}=s^{2/3}(\eta_{MN}+\delta^0_M\delta^0_N)$ we find, to quadratic
order
\begin{equation}
  S_{\text{fluid}} = \frac {Cs^{4/3}}3
   \Bigl[-3 + \frac 12 (h_{ii} + 3h_{tt})
   + \frac1{24} \bigl(-3 h_{ii}^2  + 12h_{ti}^2 - 6 \tilde h_{ij}^2 
   - 6h_{ii}h_{tt} + 9h_{tt}^2 \bigr) \Bigr] .
\end{equation}
We see a complete agreement between the expressions obtained from
holography and effective field theory.

\end{document}